Concerning the Quantum Interference Effects Associated with Electric and Magnetic Potentials [Letter Submitted to the Editor, Physics Today]

by Peter A Sturrock and Timothy R. Groves

ABSTRACT

The magnetic version of the effects commonly attributed to Aharonov and Bohm had been published ten years earlier by Ehrenberg and Siday. We compare the two articles, and conclude that their essential contents are identical. We therefore advocate that the magnetic effect be referred to as the "Ehrenberg-Siday Effect," and that the term "Aharonov-Bohm Effect" be reserved for the electrical case.

In reading "The Aharonov-Bohm effects: Variations on a subtle theme," by Herman Batelaan and Akira Tonomura (PHYSICS TODAY, September 2009, page 38), we note with some dismay that the principal (magnetic-flux) effect continues to be attributed to its re-discoverers rather than to the original discoverers. Batelaan and Tonomura write "The AB effect was already implicit in the 1926 Schrodinger equation, but it would be another three decades before physicists Yakir Aharonov and David Bohm pointed it out."[1] This is incorrect: it took only two decades before the effect was pointed out—not by Aharonov and Bohm, but by W. Ehrenberg and R.E. Siday.[2] (This prior publication was referenced by Aharonov and Bohm in their second article[3].)

According to Bohm's biographer F. David Peat, "After their first paper had been published, Bohm learned that the effect had already been postulated by a maverick physicist called Rory E. Siday."[4] It seems curious that the biographer should use the term "postulated" rather than "proposed," and that he should comment on Siday's personality but not on Ehrenberg's. One of us (PAS) knew them both: Raymond (Siday's actual name) was very smart but somewhat brash; Werner was erudite and a model of propriety, who would never have subscribed his name to an article unless he was convinced it was correct.

It seems that the main scientific (rather than sociological) reason that the AB proposal was taken more seriously than the ES proposal is that the former was presented in the context of quantum mechanics, whereas the latter was presented in the context of electron optics. A&B make a rapid transition from quantum mechanics to wave theory. E&S make a more detailed transition from geometrical optics to wave theory. Gabor's invention of holography was another interference process which was developed for application to electron optics (as well as to light optics),[5,6] and it is perhaps worth noticing that Gabor based his analysis on wave theory, not quantum mechanics.

For the magnetic case, the AB and ES articles presented identical theoretical predictions. The key (unnumbered) equation in the AB article is identical to Equation (50) in the ES article. Figure 2 of the AB article is the same as Figure 2 of the ES article. AB's estimate of the magnetic flux required to shift an interference pattern by one fringe ($4 \times 10^{-7}$ gauss cm$^2$) is to be found in the ES article. In short, the essential conclusions of the AB article had already been published in the ES article a decade earlier.

Our personal comparison is that the ES article was tedious but conceptually sound, whereas the AB article was lively but involved a conceptual leap that we find questionable. Having derived a formula for the special case of a time-varying electric field, A&B propose a relativistic generalization of their expression. However, their analysis is not Lorentz covariant, so the mention of a relativistic generalization is misleading. By contrast, E&S compute the phase shift more reliably from

$$\Delta \psi = \frac{1}{\hbar} \int \left( \mathbf{p} - \frac{e}{c} \mathbf{A} \right) \cdot d\mathbf{x}.$$

We might mention that soon after one of us (PAS) arrived at Stanford University in 1955, and before the AB publication, he asked Leonard Schiff, the pre-eminent quantum-mechanics expert at Stanford, whether Schiff found the ES proposal convincing. He did.

In the interests of historical and scholarly accuracy and of giving credit where credit is due, it would we think be appropriate to use "Ehrenberg-Siday Effect" for the magnetic interference effect, and to reserve "Aharonov-Bohm Effect" for the electrical interference effect.

We are indebted to Elliott Bloom, Dieter Kern, Walt Harrison, Peter Hawkes, Garret Moddel, Fabian Pease, Jeff Scargle, and Lenny Susskind for helpful discussions on this matter.

Peter A. Sturrock
(sturrock@stanford.edu)
Stanford University,
Stanford, California

Timothy R. Groves
(TGroves@uamail.albany.edu)
College of Nanoscale Science and Engineering
University at Albany,
State University of New York
New York